\documentclass[aps,prl,showpacs,twocolumn,preprintnumbers]{revtex4}
\usepackage{graphicx}
\begin{document}

\preprint{UMD-PP-05-052}
 \preprint{MIFP-05-30}
\title{\Large Observable $N-\bar{N}$ Oscillation in High Scale Seesaw
Models}
\author{\bf Bhaskar Dutta$^1$, Yukihiro Mimura$^1$, and R.N. Mohapatra$^2$ }

\affiliation{$^1$Dept. of Physics, Texas A\&M University,
College Station, TX 77843-4242, USA
\\
$^2$Department of Physics and Center for String and
Particle Theory, University of
Maryland, College Park, MD 20742, USA}

\date{October, 2005}

\begin{abstract}
We discuss a realistic high scale ($v_{BL}\sim 10^{12}$ GeV)
supersymmetric seesaw model based on the gauge group $SU(2)_L\times
SU(2)_R\times SU(4)_c$ where neutron-anti-neutron oscillation can be in 
the observable range without fine tuning of parameters. This is contrary 
to the naive dimensional arguments which say that $\tau_{N-\bar{N}}\propto
v_{BL}^5$ and should therefore be unobservable for seesaw scale
$v_{BL}\geq 10^{5}$ GeV. Two reasons for this enhancement are: (i) 
accidental
symmetries which keep some of the diquark Higgs masses at the weak
scale and (ii) a new supersymmetric contribution from a lower
dimensional operator. The net result is that
$\tau_{N-\bar{N}}\propto v_{BL}^2 v^3_{wk}$ rather than $v^5_{BL}$.
 The model also can explain the origin of matter via the leptogenesis 
mechanism and predicts light diquark states which can be produced at 
LHC.
\end{abstract}

\maketitle

\section{Introduction}

There are various reasons to suspect that baryon number is not a
good symmetry of nature: (i) first is that nonperturbative
effects of the standard model lead to $\Delta B\neq 0$, while
keeping $\Delta (B-L)=0$  \cite{thooft}; (ii) understanding the
origin of matter in the universe requires $\Delta B\neq
0$  \cite{sakharov} and (iii) many theories beyond the standard
model lead to interactions that violate baryon number  \cite{ps,gg}.

If indeed such interactions are there, the important question is:
can we observe them in experiments$\,$? Two interesting baryon
nonconserving processes of experimental interest are: (a) proton
decay e.g. $p\rightarrow e^++\pi^0, \bar{\nu}+K^0$ etc
\cite{gg,pdecay} and (b) $N\leftrightarrow \bar{N}$ oscillation
\cite{kuzmin,glashow,marshak}. These two classes of processes
probe two different selection rules for baryon nonconservation:
$\Delta (B-L)=0$ for proton decay and $\Delta(B-L)=2$ for
$N\leftrightarrow \bar{N}$ oscillation and indicate totally
different directions for unification beyond the standard model.
For example, observation of
 proton decay will point strongly towards a grand desert till
 about the scale of $10^{16}$ GeV whereas $N\leftrightarrow \bar{N}$
oscillation
 will require new physics at an intermediate scale at or above the
 TeV scale but much below the GUT scale.

 While proton decay goes very naturally with the idea of eventual
 grand unification of forces and matter, recent discoveries of neutrino
  oscillations have made $N\leftrightarrow \bar{N}$ oscillation to be
quite plausible theoretically if small neutrino masses are to be
understood as a consequence of the seesaw mechanism~\cite{seesaw}.
This can be seen as follows: seesaw mechanism implies Majorana
neutrinos implying the existence of $\Delta (B- L)=2$ interactions.
In the domain of baryons, it implies the existence of
$N\leftrightarrow \bar{N}$ oscillation as noted many years ago
\cite{marshak}.

An explicit model for $N\leftrightarrow \bar{N}$ oscillation
was constructed in  \cite{marshak} by implementing the seesaw
mechanism within the framework of the Pati-Salam \cite{ps}
$SU(2)_L\times SU(2)_R\times SU(4)_c$ model, where quarks and
leptons are unified. It was shown that this process is mediated by
the exchange of diquark Higgs bosons with an amplitude
 ($G_{N \leftrightarrow \bar N}$) which scales like $M^{-5}_{qq}$. In the
nonsupersymmetric version without fine tuning, one expects
$M_{qq}\propto v_{BL}$ leading to $G_{N \leftrightarrow \bar N}\simeq
v^{-5}_{BL}$. So only if $M_{qq}\sim v_{BL}\sim 10 -100$ TeV, the
  $\tau_{N\leftrightarrow \bar{N}}$ is in the range of
  $10^6-10^{8}$ sec and is accessible to experiments. On the other hand,
in generic seesaw models for neutrinos,
one expects $v_{BL}\sim 10^{11}-10^{14}$ GeV depending on whether
the third generation Dirac mass for the neutrino is 1-100 GeV. An
important question therefore is whether in realistic seesaw models,
$N\leftrightarrow \bar{N}$ oscillation is at all observable. Another
objection to the above nonsupersymmetric model for $ N \leftrightarrow
\bar N$ that was raised in the  80's was that such interactions will
erase any baryon asymmetry created at high scales. It will be therefore
important to overcome this objection.

 In this
note we point out that in a class of supersymmetric $SU(2)_L\times
SU(2)_R\times SU(4)_c$ models (called SUSY $G_{224}$), an interesting
combination of
circumstances improves the $v_{BL}$ dependence of the $G_{\Delta
B=2}$ to $v^{-2}_{BL}v^3_{wk}$ instead of $v^{-5}_{BL}$ making
$N\leftrightarrow \bar{N}$ oscillation observable. Further more the same
model also allows us to overcome the difficulty with high scale
baryogenesis. An example of such a theory was presented in  \cite{chacko}
 where it was shown that in the minimal version of the model,
  there exist accidental symmetries that imply that some of the
$M_{qq}$'s are
  in the TeV range even though $v_{BL}\simeq 10^{11}$ GeV. We
  discuss this class of theories in this letter. The new results
  in this paper are: (i) a new diagram which enhances the
$N\leftrightarrow \bar{N}$
  oscillation amplitude; (ii) a realistic example which is in
  agreement with the observed  neutrino oscillation parameters while
  predicting observable $N\leftrightarrow \bar{N}$
  oscillation amplitude and (iii) adequate leptogenesis
  mechanism \cite{fuku} for understanding the origin of matter via the
quasi-degenerate leptogenesis \cite{pilaf} with right-handed neutrinos
at $\sim 10^{7}$ GeV.

At present, the best lower bound on $\tau_{N\leftrightarrow
\bar{N}}$ comes from ILL reactor experiment \cite{milla} and is
$10^{8}$ sec. There are also comparable bounds from nucleon decay
search experiments \cite{nnbar}. There are proposals to improve the
precision of this search by at least  two orders of
magnitude \cite{kamyshkov}. We feel that the results of this paper
should give new impetus to a search for neutron-antineutron
oscillation.

\section{$SU(2)_L\times SU(2)_R\times SU(4)_c$ model with light
diquarks}

The quarks and leptons in this model are unified and transform as
$\psi:({\bf 2,1,4})\oplus \psi^c:({\bf 1,2},\bar{\bf 4})$
representations of $SU(2)_L\times SU(2)_R\times SU(4)_c$. For the
Higgs sector, we choose, $\phi_1:(\bf{2,2,1})$ and
$\phi_{15}:(\bf{2,2,15})$ to give mass to the fermions. The
$\Delta^c:({\bf 1,3,10})\oplus \bar{\Delta}^c:({\bf
1,3},\overline{\bf 10})$ to break the $B-L$ symmetry. The diquarks
mentioned above which lead to $\Delta (B-L)=2$ processes are
contained in the $\Delta^c:(\bf{1,3,10})$ multiplet. We also add a
$B-L$ neutral triplet $\Omega:(\bf{1,3,1})$ which helps to reduce
the number of light diquark states and a gauge singlet field $S$.
 The superpotential of this model is given
by: $W~=~W_Y~+~W_H$
where
\begin{eqnarray}
W_H&=&\lambda_1 S( \Delta^c\bar{\Delta}^c- M^2)~+~\lambda_A
\frac{(\Delta^c\bar{\Delta}^c)^2}{M_{P\!\ell}} \nonumber \\
&&+\lambda_B\frac{(\Delta^c{\Delta^c})(\bar{\Delta}^c\bar{\Delta}^c)}{M_{P\!\ell}}
+\lambda_C \Delta^c\bar{\Delta}^c\Omega \\
&& +\mu_{i}{\rm Tr}\,(\phi_i\phi_i)
 + \lambda_D \frac{{\rm Tr}\,(\phi_1\Delta^c  
\bar{\Delta}^c\phi_{15})}{M_{P\!\ell}} \,, \nonumber \\
W_Y&=&h_1\psi\phi_1 \psi^c + h_{15} \psi\phi_{15} \psi^c + f
\psi^c\Delta^c \psi^c.
\end{eqnarray}
Note that since we do not have parity symmetry in the model, the
Yukawa couplings $h_1$ and $h_{15}$ are not symmetric matrices.
When $\lambda_B=0$, this superpotential has an accidental global
symmetry much larger than the gauge group\cite{chacko}; as a
result, vacuum breaking of the $B-L$ symmetry leads to the
existence of light diquark states that mediate $N\leftrightarrow
\bar{N}$ oscillation and enhance the amplitude. In fact it was
shown that for $\langle\Delta^c\rangle\sim
\langle\bar{\Delta}^c\rangle\neq 0$ and $\langle\Omega\rangle\neq
0$ and all VEVs in the range of $10^{11}$ GeV, the light states
are those with quantum numbers: $\Delta_{u^cu^c}$.
 The symmetry argument behind is that \cite{chacko}
 for $\lambda_B=0$, the above superpotential is invariant under
$U(10,c)\times SU(2,c)$ symmetry  which breaks down to
$U(9,c)\times U(1)$ when $\langle\Delta^c_{\nu^c\nu^c}\rangle =
v_{BL}\neq 0$. This results in 21 complex massless states; on the
other hand these vevs also breaks the gauge symmetry down from
$SU(2)_R\times SU(4)_c$ to $SU(3)_c\times U(1)_Y$. This allows
nine of the above states to pick up masses of order $gv_{BL}$
leaving 12 massless complex states which are the six
$\Delta^c_{u^cu^c}$ plus six $\bar{\Delta}^c_{u^cu^c}$ states.
Once $\lambda_B\neq 0$, they pick up mass (call $M_{u^cu^c}$) of
order of the elctroweak scale.

\bigskip
\section{ $N\leftrightarrow \bar{N}$ oscillation- a new diagram}
\medskip

To discuss $N\leftrightarrow \bar{N}$ oscillation,
we introduce a new term in the superpotential of the form:
\begin{equation}
W_{\Delta B=2}=\frac{1}{M_*}\epsilon^{\mu'\nu'\lambda'\sigma'}
\epsilon^{\mu\nu\lambda\sigma}
\Delta^c_{\mu\mu'}\Delta^c_{\nu\nu'}\Delta^c_{\lambda\lambda'}
\Delta^c_{\sigma\sigma'}\,,
\end{equation}
where the $\mu,\nu  $ etc stand for $SU(4)_c$ indices and we have
suppressed the $SU(2)_R$ indices. We choose $M_*\ll M_{P\!\ell}$. This does 
not affect the masses
of the Higgs fields. When $\Delta_{\nu^c\nu^c}^c$ aquires a VEV,
$\Delta B = 2$ interaction are induced from the superpotential,
%
The contribution to neutron antineutron oscillation in this model comes 
from the diagram in Fig. 1, which gives
\begin{eqnarray}
G_{N \leftrightarrow \bar N}\simeq\frac{g^2_3}{16\pi^2}
\frac{f^3_{11}v_{BL}}{M^2_{u^cu^c}M^2_{d^cd^c}M_{\rm SUSY}M_*},
\end{eqnarray}
which scales like $v^{-2}_{BL} v_{wk}^{-3}$. There is also another 
pure Higgs diagram that involves the vertex 
$\tilde{d^c}\tilde{d^c}\Delta_{d^cd^c}\Delta_{u^cu^c}$ which also 
scales the same way and gives 
an identical contribution. This graph is the supersymmetric partner of 
Fig. 1. Note that for high scale seesaw models, these contributions to 
$G_{N \leftrightarrow \bar N}$ 
are considerably enhanced over that from the nonsupersymmetric 
theory\cite{marshak} which goes like $v^{-5}_{BL}$. 

 \begin{figure}[tbp]
 \includegraphics[width=6cm]{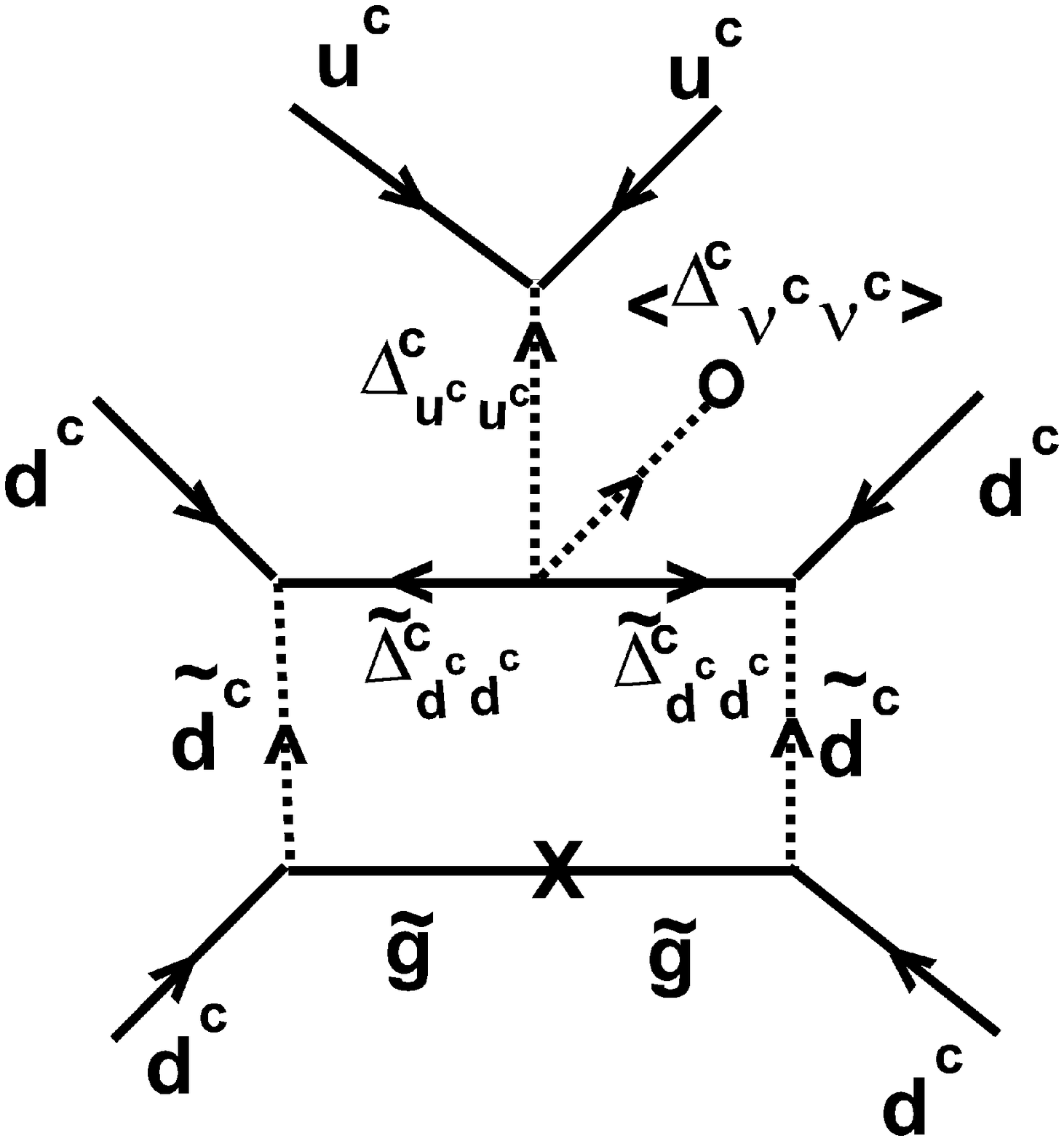}%
 \caption{The new Feynman diagram for $N-\bar{N}$ oscillation.}
 \end{figure}

In order to estimate the rate for $N\leftrightarrow \bar{N}$
oscillation, we need not only the different mass values for which we
now have an order of magnitude, we also need the Yukawa coupling
$f_{11}$. Now $f_{11}$ is a small number since its value is associated
with the lightest right-handed neutrino mass. However, in the calculation
we need its value in the basis where quark masses are diagonal. We
note that the
 $N-\bar{N}$ diagrams involve only the
right-handed quarks, the rotation matrix need not be the CKM
matrix. The right-handed rotations need to be large e.g. in order
to involve $f_{33}$ (which is $O(1)$), we need
$(V_R^{(u,d)})_{31}$ to be large, where $V_L^{(u,d)\dagger}
Y_{u,d}V_R^{(u,d)}=Y_{u,d}^{\rm diag.}$. The left-handed rotation
matrices $V_L^{(u,d)}$ contribute to the CKM matrix, but
right-handed rotation matrices $V_R^{(u,d)}$ are unphysical in the
standard model. In this model, however, we get to see its
contribution since we have a left-right gauge symmetry as we will
see later.

Let us now estimate the time of oscillation. When we start on a
$f$-diagonal basis (call the diagonal matrix $\hat f$), the Majorana
coupling $f_{11}$ in the diagonal basis of up- and down-type quark
matrices can be written as $(V_R^T {\hat f} V_R)_{11} \sim
(V^R_{31})^2 \hat f_{33}$. Now $\hat f_{33}$ is $O(1)$ and
$V^R_{31}$ can be $\sim 0.6$, so $f_{11}$ is about 0.4 in the
diagonal basis of the quark matrices. We use $M_{\rm SUSY}$,
$M_{u^cu^c}$ $\sim 350$ GeV and $v_{BL}\sim 10^{12}$ GeV. Due to the
presence of the colored field the couplings become non-perturbative
very soon beyond the $v_{BL}$, we therefore take $M_{*}\simeq
10^{13}$ GeV. The mass of $\tilde\Delta_{d^cd^c}$ i.e. $M_{
d^cd^c}$ is $10^{9}$ GeV which is obtained from the VEV of
$\Omega:(\bf{1,3,1})$. Putting all the above the numbers together, we get
$G_{N \leftrightarrow \bar N}\simeq 1 \cdot 10^{-30} \ {\rm GeV^{-5}}$.
Along with the hadronic matrix element\cite{Rao}, the
$N-\bar{N}$ oscillation time is found to be about $2.5\times
10^{10}$ sec which is within the reach of possible next generation
measurements.

We also note that as noted in \cite{marshak}
the model is invariant under the hidden discrete symmetry under which a field
$X \rightarrow e^{i\pi B_{X}}X$, where $B_X$ is the baryon number of the
field $X$. As a result, proton is absolutely stable in the model.

%


\bigskip
\section{ Leptogenesis and $N\leftrightarrow{\bar N}$}
\medskip

Since $N\leftrightarrow \bar{N}$ oscillation is a low intermediate
scale phenomenon, it is necessary to see how it affects the
baryogenesis discussion. One could of course invoke weak scale
baryogenesis, which would remain unaffected since it happen at the
weak scale. Instead here we focus on leptogenesis mechanism for
origin of matter. The obvious problem here is that if leptogenesis
takes place at a scale where the 
$\Delta B=2$ process is in thermal equilibrium, then the
matter-antimatter asymmetry generated will be erased. So two questions
must be answered within our model: (i) what is the temperature $T_D$
below which the $\Delta B=2$ processes are out of equilibrium$\,$? and
(ii) what constraints must be obeyed by the right-handed neutrino
spectrum in the theory so that it will be possible to generate the
lepton asymmetry below $T_D$? Clearly, $T_D$ must be above the weak
scale so that the sphalerons can be effective in converting leptons
to baryons.

To answer the first question, we compare the $\Delta B=2$ reaction
rate with the Hubble expansion.
The diquark Higgs field $\Delta^c_{u^cu^c}$ remains at weak scale in the thermal bath,
while $\Delta_{d^cd^c}$ etc decouple at high scale.
The $\Delta B=2$ process surviving till lowest temperature is
$\Delta_{u^cu^c}^c \leftrightarrow b^c b^c \tilde b^c \tilde b^c$.
The coupling of the interaction can be estimated as
$G_{\Delta B=2} =
\frac{v_{BL}}{M^2_{d^cd^c} M_*} \simeq 10^{-19} \ {\rm GeV}^{-2}$.
%
%
%
The out of equilibrium condition of this process would be:
$G_{\Delta B=2}^2 T^{5} \alt g^{1/2}_*\frac{T^2}{M_P} \,$
which gives $T_D \simeq 10^{7}$ GeV. This implies
that for leptogenesis to work, at least one right-handed
 neutrino should be lighter than $10^7$ GeV. In fact in view of
the low scale, we will need to invoke quasi-degenerate
leptogenesis \cite{pilaf}, in the case of thermal leptogenesis or
one can also have the inflaton decaying into the two light
right-handed neutrinos and the decay of the right-handed neutrinos
would generate the required lepton asymmetry. In both these cases,
the lighter right-handed neutrinos are almost degenerate to produce the
correct amount of baryon asymmetry. Below, we give an example of
the heavy right-handed neutrino as well as the Dirac neutrino mass
texture as an example that leads to the correct neutrino mixings
as well as a correct amount of baryons.
 We choose the Majorana neutrino coupling to the $\Delta$s as:
\begin{equation}
f=\left(\begin{array}{ccc}
0 & x & 0 \\
x & 0 & 0 \\
0 & 0 & 1\end{array} \right),
\end{equation}
where $x\sim 10^{-5}$ to
have two right-handed neutrino around $10^7$ GeV, where
 the heaviest neutrino mass is $10^{12}$ GeV.
The degeneracy has to be broken, however, since
 we need $|M_{N_1}-M_{N_2}|\geq\Gamma_{N_1}+\Gamma_{N_2}$,
where $M_{N_i}$ and $\Gamma_{N_i}$ are masses and decay widths of
right-handed neutrinos, respectively.
Thus, we need small perturbation in 11 and 22 elements.
 The formula for lepton asymmetry is:
\begin{eqnarray}
\epsilon_i \simeq \frac{1}{8\pi(\lambda_\nu^D{}^\dagger \lambda_\nu^D)_{ii}} \sum_{k\neq i}
{\rm Im} (\lambda_\nu^D{}^\dagger \lambda_\nu^D)^2_{ik}F\left(\frac{M_{N_k}}{M_{N_i}}\right),
\end{eqnarray}
where the neutrino Dirac Yukawa coupling $\lambda_\nu^D$ in this
expression is given in the basis where the right-handed Majorana
neutrino mass matrix is diagonal. We choose
 the following Dirac neutrino
coupling
(which generates the correct neutrino masses using type I
seesaw)
$$\lambda_\nu^D=\left(\begin{array}{ccc}
1\times10^{-6} & 3\times10^{-5}  & 0.026 \\
5\times10^{-6} & 5\times 10^{-6}  & 0.026 \\
0.00031& 0.00031 & -0.013\end{array} \right),$$
where we have omitted phases of order one.
The asymmetry generated by this texture needs to be multiplied by
the efficiency factor $\kappa_i$. For quasi degenerate
right-handed neutrinos, one gets $F(\frac{M_{N_2}}{M_{N_1}})\simeq
\frac{M_{N_1}}{M_{N_1}-M_{N_2}}$. As a result, there is an
enhancement factor for the lepton asymmetry depending on the mass
difference which is a parameter in this model.
 The decay
parameter $K_i(\equiv\Gamma_i/2H)$, where $\Gamma_i$ is the decay
width of the heavy Majorana neutrino, is large; for example, $K_i
\propto (\lambda_\nu^D{}^\dagger \lambda_\nu^D)_{ii}/M_{N_i}$ for
above choice of $\lambda^D$ and $M_{N_{1,2}}$ around $10^7$ GeV,
$K_i\simeq 10^2$ or smaller.

 Our model has several interesting phenomenological
consequences both for colliders and rare decays for up-type quarks. 
For instance, since our diquark Higgs bosons and
their superpartners have masses in the few hundred GeV range, 
they should be produced at LHC, possibly at Tevatron. The
production can happen singly or in pairs via parton processes such as 
$\bar{q}+\bar{q} \rightarrow \Delta^c_{u^cu^c}$,
$q+\bar{q}\rightarrow \Delta^c_{u^cu^c}{\Delta}^c{}^*_{u^cu^c}$ etc. 
These new Higgs boson and its superpartners can
decay into to two charm, up quarks, top quarks should provide a good signal. 
At the LHC, the signal from a pair
production may contain the following final states: 4 charm quarks,
 2 top-2 charm quarks, 4 top quarks etc. The fermionic
 partner of these diquark Higgs, however, would produce missing energy in 
the final states. The single production will
 produce a change in the top quark pair production. A detailed study is 
underway.
 A second interesting consequence is
that in the mass eigenstate basis for quarks, the light
$\Delta_{u^cu^c}^c$ will mediate flavor changing processes such as
$D\rightarrow\pi\pi$, $t\rightarrow c+ G$ etc. The present bounds on
charm changing processes is
consistent with the value of the diquark Higgs boson mass that has
been used in the paper.

\bigskip
\section{ Conclusion}
\medskip

In conclusion, we have presented a quark-lepton unified model where
despite the high seesaw ($v_{BL}$) scale in the range of $\sim
10^{12}$ GeV, the $N-\bar{N}$ oscillation time is around
$10^{10}$ sec. due to the presence of a new supersymmetric
graph and accidental symmetries of the Higgs potential (also
connected to supersymmetry). No unatural fine tuning of diquark
masses is needed to get such an enhanced effect. Our predicted
oscillation time is within the reach of possible future experiments.
We believe that this work should provide a new motivation to conduct
a new round of search for $N-\bar{N}$ oscillation. To show that this
is a completely realistic model, we note that (i) the proton is
absolutely stable in this model; (ii) 
the model can provide a realistic description of
known quark and lepton physics as well as (iii) a way to generate an
adequate lepton asymmetry via the mechanism of 
quasi-degenerate leptogenesis.

 This work of R. N. M. is supported by the National
Science Foundation grant no. Phy-0354401. R. N. M. would like to thank S. 
Nasri for comments.

\end{document}